\documentstyle[amstex,art8,aas2pp4,flushrt,tighten,xr,epsfig]{article}

\newcommand{\pa}{\partial}

\newcommand{\vth}{v_{\rm th}}
\newcommand{\dis}{\displaystyle}

\newcommand{\cri}{_{\rm c}}
\newcommand{\ze}{z_{\rm e}}
\newcommand{\gl}{g_{\rm l}}
\newcommand{\vecn}{{\bf n}}
\newcommand{\vecv}{{\bf v}}
 
\received{}
\accepted{}
\journalid{}{}
\articleid{}{}

\slugcomment{Submitted to the Astrophysical Journal}

\begin{document}

\title{ABBOTT WAVE-TRIGGERED RUNAWAY IN LINE-DRIVEN WINDS FROM STARS
   AND ACCRETION DISKS}

\author{Achim Feldmeier}

\affil{Astrophysik, Institut f\"ur Physik, Universit\"at Potsdam, Am
Neuen Palais 10, 14469 Potsdam, Germany, E-mail: {\tt
afeld@@astro.physik.uni-potsdam.de}}

\and

\author{Isaac Shlosman\altaffilmark{1,2}}

\affil{Joint Institute for Laboratory Astrophysics, University of Colorado,
    Box 440, Boulder, CO 80309-440, USA, E-mail: {\tt shlosman@@pa.uky.edu}}

\altaffiltext{1}{JILA Visiting Fellow}
\altaffiltext{2}{permanent address: Department of Physics and Astronomy, 
University of Kentucky, Lexington, KY 40506-0055}

\begin{abstract}

Line-driven winds from stars and accretion disks are accelerated by
scattering in numerous line transitions. The wind is believed to adopt
a unique critical solution, out of the infinite variety of shallow and
steep solutions. We study the inherent dynamics of the transition
towards the critical wind. A new runaway wind mechanism is analyzed in
terms of radiative-acoustic (Abbott) waves which are responsible for
shaping the wind velocity law and fixing the mass loss. Three
different flow types result, depending on the location of
perturbations. First, if the shallow solution is perturbed
sufficiently far downstream, a single critical point forms in the
flow, which is a barrier for Abbott waves, and the solution tends to
the critical one. Second, if the shallow solution is perturbed
upstream from this critical point, mass overloading results, and the
critical point is shifted inwards. This wind exhibits a broad,
stationary region of decelerating flow and its velocity law has kinks.
Third, for perturbations even further upstream, the overloaded wind
becomes time-dependent, and develops shocks and dense shells.

\end{abstract}


\section{Introduction}
\label{introduction}

Radiation-driven winds which are accelerated by absorption and
re-emission of continuum photons in spectral lines form an interesting
class of hydrodynamic flows, termed line-driven winds (LDWs). They
occur in OB and W-R stars, cataclysmic variables, and probably in
active galactic nuclei and luminous young stellar objects. These winds
are characterized by a unique dependence of line force on the velocity
gradient in the flow. This causes a new, radiative wave type.

The nature of these waves was first discussed by Abbott (1980,
hereafter `Abbott waves'), who found two modes, a slow, acoustic one
propagating downstream, and a fast, radiative one propagating
upstream. The critical point found by Castor, Abbott, \& Klein (1975,
hereafter CAK) from analysis of the stationary Euler equation is a
barrier for these waves, the same way as the sonic point is to sound
waves. The intriguing property of Abbott waves is that they propagate
downstream slower than the sound speed, while they propagate upstream
at very large speeds, highly supersonically. The latter fact reflects
essentially the radiative nature of Abbott waves.

No such upstream propagating, radiative mode was found by Owocki \&
Rybicki (1986), who calculated the Green's function for winds driven
by pure line absorption. The explanation is that the pure absorption
case suppresses the radiative upstream mode, as photons can propagate
only downstream. The radiative upstream mode is back when line
scattering is included (Owocki \& Puls 1999).

The question arises for the physical interpretation of Abbott waves.
Do they represent a physical entity which is responsible for shaping
the flow by communicating essential flow properties between different
points in the wind? In particular, could Abbott waves be the prime
cause for evolution of LDWs towards a CAK-type, steady-state solution?

The analysis by CAK of the steady-state Euler equation for LDWs has
revealed an infinite family of mathematical solutions, but only one,
hereafter `critical solution,' which extends from the photosphere to
arbitrary large radii. Other solutions either do not reach infinity or
the photosphere. The former solutions are called shallow and the
latter ones --- steep. The unique, critical wind starts as the fastest
shallow solution and switches smoothly to the slowest steep solution
at the critical point.

The shallow wind solutions found by CAK are the analog to solar wind
breezes, in that they are sub-abbottic everywhere. They were abandoned
by CAK because they cannot provide the required spherical expansion
work at large radii. This exclusion of shallow solutions can be
criticized in different respects: (i) the breakdown happens only
around 300 stellar radii, where basic assumptions of the model (fluid
description; spherical symmetry; isothermality) may become invalid.
(ii) Shallow solutions could be extended to infinity by jumping at
some large radius to a decelerating wind branch. The latter was
excluded a priori by CAK. However, this jump can occur beyond a few
stellar radii, where the wind has already reached its local escape
speed. (iii) For models of disk LDW, even the critical solution itself
does not extend to infinity, and becomes imaginary beyond a certain
radius (Feldmeier \& Shlosman 1999). Jumps to the decelerating branch
are unavoidable then.

With shallow solutions being valid stationary solutions, the question
arises, what forces the wind to adopt the critical CAK solution?
Numerical aspects of this question have been discussed by Feldmeier,
Shlosman \& Hamann (2001), who noted that outer boundary conditions
and a Courant time step which do not account for Abbott waves can set
off numerical runaway, often towards the critical solution.

In the present paper, we focus on a physical interpretation of Abbott
waves, extending our previous work on this subject (Feldmeier \&
Shlosman 2000). We show that these waves are the prime driver of
evolution of LDWs towards a unique, steady-state solution
characterized by a specific velocity law and mass loss rate. In
particular, we find that as is the case for solar wind breezes,
shallow solutions can evolve due to waves which propagate upstream to
the wind base (the photosphere). As a new effect in LDWs, Abbott waves
`drag' the solution in one preferred direction, towards larger
velocities. The wind becomes stable when a critical point forms,
through which outer perturbations can no longer penetrate inwards.

\section{Abbott waves}

\subsection{Wind model}
\label{windmodel}

Only wind acceleration due to a line force in Sobolev approximation
for radiative transfer is considered in this paper. The large number
of lines driving the wind is dealt with using a CAK line distribution
function. The latter is characterized by a power law index $\alpha$,
which lies between 0 and 1. The Sobolev force is proportional to
 \begin{equation}
 \label{genlifo}
 \gl \sim \int d\omega \,\vecn I_\vecn \tau_\vecn^{-\alpha},
 \end{equation}
 where $\vecn$ is the unit vector pointing in the direction of the
surface angle element $d\omega$, $I$ is the frequency integrated
specific intensity, and $\tau$ is the Sobolev optical depth,
 \begin{equation}
 \tau_\vecn = {\kappa\rho\vth \over \vecn \cdot(\vecn \cdot(\nabla
\vecv))},
 \end{equation}
 with rate-of-strain tensor $\nabla\vecv$, mass absorption coefficient
$\kappa$, and ion thermal speed $\vth$. The force proportionality
constant is fixed later in terms of the critical solution. Through
$\nabla\vecv$, the line force depends on the streamline geometry. To
simplify, we take $\tau$ in (\ref{genlifo}) out of the integral and
replace it by an average or {\it equivalent} optical depth in one
direction. Specifically, the latter is assumed to be the flow
direction. This approach corresponds to the CAK `radial streaming
approximation.' For a planar flow with height coordinate $z$, the line
force becomes, assuming constant $\kappa$ and $\vth$,
 \begin{equation}
 \gl \sim F (v'/\rho)^\alpha,
 \end{equation}
 where $v'=\pa v/\pa z$, and the radiative flux $F$ is a function of
$z$. Even this highly idealized line force depends in a non-linear way
on the hydrodynamic variables $v'$ and $\rho$.

So far, the following assumptions were introduced: (i) the line force
is calculated in Sobolev approximation, using (ii) a CAK line
distribution function (without line overlap), and (iii) applying the
radial streaming approximation for (iv) planar flow. To these, we add
the further assumptions: (v) the flux $F$ is constant with $z$, but
gravity $g$ may depend arbitrarily on $z$. This can serve to model
winds from thin, isothermal accretion disks, in which case $g$ grows
first linearly with $z$ and at large distances drops off as $z^{-2}$.
Alternatively, with $F$ {\it and} $g$ being constants, one can model
the launch region of a stellar wind, but a well-known degeneracy
occurs here (Poe, Owocki, \& Castor 1990). (vi) Zero sound speed is
assumed, $a=0$, and (vii) we fix $\alpha=1/2$. Note that the Sobolev
line force is independent of $\vth$ and therefore of $a$. LDWs are
hypersonic, and except near the photosphere, gas pressure plays no
role. The 1-D continuity and Euler equations are,
\begin{gather}
 \label{continuity}
 {\pa \rho \over \pa t} + v {\pa \rho \over \pa z} + \rho {\pa v
 \over \pa z} = 0,\\
 \label{euler} 
 E \equiv {\pa v \over \pa t} + v\, {\pa v \over \pa z} + g(z) -
 C_0 \, F \sqrt{{\pa v / \pa z \over \rho}}=0,
 \end{gather}
 with constant $C_0$. We consider first stationary solutions, $\rho
v={\rm const}$. A normalized quantity $m=\rho v/\rho\cri v\cri$ is
introduced, where $\rho\cri(z)$ and $v\cri(z)$ are the density and
velocity law of the critical wind, which is defined below. Besides
$m$, a second, new hydrodynamic variable, $w'=vv'$, is defined, and
the Euler equation becomes,
 \begin{equation}
 \label{eulerstat}
 w' + g(z) - C\sqrt{w'/m} =0.
 \end{equation}
 The flux $F$ was absorbed into the constant $C$. At each $z$,
(\ref{eulerstat}) is a quadratic equation in $\sqrt{w'}$, with
solutions,
 \begin{equation}
 \label{eulerstatsolu}
 \sqrt{w'} = {1\over 2\sqrt{m}} \left(C \pm \sqrt{C^2-4gm}\right).
 \end{equation}
 The velocity law $v(z)$ is obtained from $w'$ by quadrature.
Solutions with `$-$' are termed {\it shallow}, those with `$+$' are
termed {\it steep}. If $m<1$ (see below), shallow and steep solutions
exist from $z=0$ to $\infty$. If $m>1$, shallow and steep solutions
become imaginary in a certain $z$ interval. In this region, the line
force $\sim C/\sqrt{m}$ cannot balance gravity $g$.

Within the family of shallow wind solutions, the mass flux $\rho v$
increases monotonically with terminal speed, while for steep solutions
the trend is opposite. The largest mass flux which keeps the solution
everywhere real defines the critical wind, $m\cri=1$. Setting the
square root in eq.~(\ref{eulerstatsolu}) to 0, this implies for $C$,
 \begin{equation}
 \label{constantC}
 C = 2\sqrt{g\cri},
 \end{equation}
 where $g\cri=g(z\cri)$ means gravity at the critical point of the
critical solution. How is $z\cri$ found? Differentiating the
stationary Euler equation, $E\bigl(z,w'(z)\bigr)=0$, with respect to
$z$ and using $\pa E/ \pa w'\cri=0$ at the critical point (crossing of
solutions), one finds
 \begin{equation}
 0={dE\over dz\cri} = {dg\over dz\cri}.
 \end{equation}
 Hence, the critical point coincides with the gravity maximum. This is
not an accident, but expresses that the critical point lies at the
bottleneck of the flow, as for a Laval nozzle (Abbott 1980). If the
flux $F$ varies with $z$, the generalized area function depends also
on $F$, and the critical point no longer coincides with the gravity
maximum (see Feldmeier \& Shlosman 1999 for examples). If $g={\rm
const}$, the critical point degenerates, and every point in the flow
becomes critical. In stellar wind calculations, the correct critical
point location is found by including the finite cone correction factor
for the stellar disk as an `area' function (Pauldrach, Puls, \&
Kudritzki~1986; Friend \& Abbott 1986).

The wind solution becomes,
 \begin{equation}
 \label{cakwindsol}
 w'={g_c \over m} \left( 1\pm \sqrt{1-{mg \over g_c}}\right)^2.
 \end{equation}
 At the critical point, shallow and steep solutions with $m=1$ merge
in such a way that the slope in passing from one to the other is
continuous. Staying instead on either shallow or steep solutions
introduces a discontinuity in $v''(z\cri)$. Discontinuities in
derivatives of hydrodynamic variables, termed weak discontinuities,
lie on flow characteristics (Courant \& Hilbert 1968). Characteristics
are the space-time trajectories of wave phases. Indeed, we find below
that the critical point is a barrier for Abbott waves.

It is at this point that the question arises, which solution the wind
adopts --- a shallow, steep, or critical one. This issue will be
resolved by discussing {\it runaway} of shallow solutions. We shall
find that Abbott waves are the prime driver of this evolution.

\subsection{Green's function}

We derive the Green's function for Abbott waves in Sobolev
approximation. The Green's function gives the response of a medium to
a localized, delta function perturbation in space and time, and is
complementary to the harmonic dispersion analysis of Abbott (1980) and
Owocki \& Rybicki (1984). Since localized perturbations contain many
harmonics, a Green's function describes wave interference. This is
clearly seen for surface water waves, whose Green's function is known
from Fresnel diffraction in optics (Lamb 1932, p.~386). For
simplicity, we consider only a single, optically thick line, with
Sobolev force ($\alpha\equiv 1$),
 \begin{equation}
 \gl=A\; {\pa v\over\pa z}.
 \end{equation}
 Density $\rho$ was absorbed into the constant $A$. We assume WKB
approximation to hold (slowly varying background flow) and consider
velocity perturbations only. The characteristic analysis in the next
section will show that the Abbott wave amplitude is $v'/\rho$, hence,
Abbott waves are not annihilated by this restriction to velocity
perturbations. The linearized Euler equation for small perturbations
is
 \begin{equation}
 \label{eq1}
 {\pa\over\pa t} \delta v(z,t) = \delta \gl(z,t)= A\, \delta v'(z,t).
 \end{equation}
 The Green's function problem is posed by specifying as initial
conditions,
 \begin{equation}
 \label{eq2}
 \delta v(z,0)=\delta(z-z_0).
 \end{equation}
 Multiplying eq.~(\ref{eq1}) by $e^{-ikz}$ and integrating over $z$,
 \begin{equation}
 \label{eq3}
 {\pa\over\pa t} \overline{\delta v}(k,t) = ikA \overline{\delta
v}(k,t),
 \end{equation}
 where a bar indicates Fourier transforms, $\overline{\delta v}=$
$\int \delta v$ $e^{-ikz} dz$. The right hand side was obtained by
integration by parts, assuming $\delta v(-\infty,t)= \delta
v(\infty,t)=0$.  This is shown a posteriori. The solution of
(\ref{eq3}) is
 \begin{equation}
 \overline{\delta v}(k,t)=be^{ikAt},
 \end{equation}
 with constant $b$. Fourier transforming (\ref{eq2}) from $z$ to $k$
space,
 \begin{equation}
 \overline{\delta v}(k,0)=e^{-ikz_0}=b,
 \end{equation}
 and
 \begin{equation}
 \overline{\delta v}(k,t)=e^{ik(At-z_0)}.
 \end{equation}
 Fourier transforming back to $z$ space,
 \begin{equation}
 \label{eq73}
 \delta v(z,t)= {1\over 2\pi} \int_{-\infty}^\infty dk \, e^{ikz}
e^{ik(At-z_0)}= \delta(z-z_0+At).
 \end{equation}
 Therefore, the initial delta function propagates without dispersion
towards smaller $z$, at an Abbott speed $-A$. Furthermore, $\delta
v=0$ at $z=\pm\infty$, as assumed. Since no wave dispersion occurs,
the same Abbott speed $A$ is also obtained by considering {\it
harmonic} perturbations. Inserting $\delta v=\overline{\delta v} \,
e^{i(kz-\omega t)}$ in (\ref{eq1}) gives as phase and group speed,
 \begin{equation}
 {\omega\over k} = {d\omega\over dk} = -A.
 \end{equation}
 The Green's function, $G$, is defined by ($F$ an arbitrary function),
 \begin{equation}
 \label{defgreen}
 F(z,t)=\int_{-\infty}^\infty dz' \,G(z-z',t) \,F(z',0).
 \end{equation}
 From (\ref{eq2}) and (\ref{eq73}),
 \begin{equation}
 G(z,t)=\delta(z+At),
 \end{equation}
 a result first obtained by Owocki \& Rybicki (1986).

The present case of optically thick lines only corresponds to
$\alpha=1$. An explicit expression for the Abbott speed is not
relevant then: opposed to all cases $\alpha<1$, $\alpha=1$ poses {\it
no} eigenvalue problem for $m$. We return therefore to $\alpha=1/2$.

\subsection{Abbott wave characteristics}

Besides a harmonic and Green's function analysis, a characteristic
analysis can be given for Abbott waves. Especially, the latter is not
restricted to linear waves. Inserting $C$ from (\ref{constantC}), the
equations of motion (\ref{continuity}, \ref{euler}) become (dots
indicate time derivatives),
 \begin{gather}
 \label{confun}
 \dot \rho + v\rho' + \rho v' =0,\\
 \label{eulfun}
 \dot v + v\,v' + g(z) - 2\Gamma \sqrt{v' \over \rho} =0,
 \end{gather}
 where we introduced the constant
 \begin{equation}
 \Gamma=\sqrt{g\cri \rho\cri v\cri}.
 \end{equation}
 
 To bring these equations into characteristic form, we first write the
continuity equation formally as $K(\rho,v)=0$, the Euler equation as
$E(\rho,v)=0$. For non-linear, first order systems of partial
differential equations, $K=0$ and $E=0$, in two unknown variables
$\rho$ and $v$, the latter being functions of coordinates $t$ and $z$,
the characteristic directions or speeds, $a$, are determined by
(Courant \& Hilbert 1968; vol.~II, chap.~5, \S 3)
 \begin{equation}
 \label{vvmatrix}
 \begin{vmatrix}
 -a K_{\dot \rho} + K_{\rho'} & -a K_{\dot v} + K_{v'} \\
 -a E_{\dot \rho} + E_{\rho'} & -a E_{\dot v} + E_{v'} 
 \end{vmatrix}
 =0,
 \end{equation}
 where $K_{\rho'}=\pa K/\pa\rho'$, $E_{\dot v}=\pa E/\pa\dot v$, etc.
We use the symbol $a$, hitherto reserved for the sound speed, also for
characteristic speeds. The meaning should be clear from the
context. Inserting $K$ and $E$ in (\ref{vvmatrix}),
 \begin{equation}
 \begin{vmatrix}
 -a+v & \rho \\
 0    & -a+v-{\displaystyle \Gamma\over \displaystyle\sqrt{\rho v'}}
 \end{vmatrix}
 =0,
 \end{equation}
 hence,
 \begin{equation}
 \label{abtsp}
 a_+ = v,\qquad a_- \equiv A = v-{\Gamma\over\sqrt{\rho v'}},
 \end{equation}
 in the observers frame. The Abbott speed is again denoted $A$. In the
comoving frame, $a_+ = 0$ and $a_- = -\Gamma/\sqrt{\rho v'}$. The
downstream ($+$), slow wave mode corresponds to sound waves. The
upstream ($-$), fast mode is of radiative origin.

A simple, heuristic argument can be given for the occurence of Abbott
waves. Consider a long scale perturbation of a stationary velocity
law. At the node where the velocity gradient gets steepened, the
Sobolev line force increases. The gas is accelerated to larger speeds,
hence the node shifts inwards. Similarly, the node where the velocity
law becomes shallower shifts inwards. The node shift corresponds to
phase propagation of a harmonic wave.

Next, we bring the equations of motion into characteristic form. To
this end, the Euler equation is quasi-linearized by differentiating it
with respect to $z$ (Courant \& Hilbert 1968), introducing a new,
fundamental variable $f=v'$,
 \begin{equation}
 \dot f + vf' + f^2 - {\Gamma \over\sqrt{f\rho}}
 \left( f'-f{\rho' \over \rho}\right) +g'=0.
 \end{equation}
 Re-bracketing and multiplying with $\rho$,
 \begin{equation}
 \rho\dot f + \rho Af' + \rho f^2 + {\Gamma \over\sqrt{f\rho}} f\rho' +
\rho g'=0.
 \end{equation}
 Using the continuity equation,
 \begin{align}
 0&= \rho\dot f + \rho Af' - f\dot\rho - \left(v- {\Gamma
\over\sqrt{f\rho}}\right) f\rho' +\rho g'\notag\\
  &= \rho\dot f + \rho Af' - f\dot\rho - A f\rho' +\rho g'\\
  &= \rho^2 \left(\pa_t + A \pa_z \right){f\over \rho}+\rho g'.\notag
 \end{align}
 The Euler equation in characteristic form is therefore,
 \begin{equation}
 \left(\pa_t + A \pa_z \right) {v'\over \rho}= -{g'\over \rho},
 \end{equation}
 with $A$ from (\ref{abtsp}). We assume that WKB approximation
applies, i.e., that the temporal and spatial derivatives on the left
hand side are individually much larger than the right hand side, hence
the latter can be neglected. In a frame moving at speed $-A$, the
function $v'/\rho$ is constant, and can be interpreted as a wave
amplitude. Note that $v'/\rho$ is inversely proportional to the
Sobolev line optical depth, indicating that Abbott waves are indeed a
radiative mode.

Introducing $f$ in the continuity equation puts it into characteristic
form,
 \begin{equation}
  \left(\pa_t + v \pa_z \right) \rho =-f\rho.
 \end{equation}
 Here, $f\rho$ is an inhomogeneous term. WKB approximation cannot be
assumed here, since $f$ may vary on short scales. The wave amplitude,
$\rho$, is no longer constant along $v$ characteristics, but changes
according to this (ordinary) differential equation. Since gas pressure
$p$ scales with density, this equation shows that the outward mode
corresponds to sound.

For {\it stationary} winds, the Abbott speed in the observers frame
becomes ($\alpha=1/2$),
 \begin{equation}
 \label{staabbspee}
  A=v \left(1-\sqrt{g_c \over mw'} \right)
   =v \left(1-{1\over\sqrt{mw'/w'_c}} \right),
 \end{equation}
 since $w\cri'=g\cri$ from (\ref{eulerstatsolu}, \ref{constantC}).
With $A\cri=0$, the critical point is a stagnation point for Abbott
waves. For shallow winds, $m<1$ and $w'/w'_c<1$, hence $A<0$ and
Abbott waves propagate upstream from any $z$ to the photosphere
located at $z=0$. Shallow solutions are, therefore, sub-abbottic, and
are the analog to solar wind breezes. For steep solutions,
$\sqrt{mw'/w'_c}>1$ from (\ref{cakwindsol}). Hence $A>0$, and steep
solutions are super-abbottic. Once the wind has adopted a steep
solution, the flow can no longer communicate with the wind base: steep
solutions cannot evolve by means of Abbott waves.

\subsection{Negative velocity gradients}

A surprising result occurs when we allow for {\it negative} velocity
gradients, $v'<0$, somewhere in the wind. This corresponds to flow
deceleration, not necessarily to accretion instead of wind. The
Sobolev force is blind to the sign of the velocity gradient. All that
is important is the presence of a velocity gradient, to Doppler-shift
ions out of the absorption shadow of intervening ions. Hence, a
natural generalization of the Sobolev line force is,
 \begin{equation}
 \gl =  2\Gamma \sqrt{|v'| \over \rho}.
 \end{equation}
 This holds for a purely local force. However, if $v'<0$, the velocity
law is non-monotonic and multiple resonance locations occur. Radiative
transfer is no longer local, as photons are absorbed and scattered at
different locations. The incident flux is no longer determined by the
photospheric flux $F$ alone, but forward and backward scattering has
to be accounted for. The constant $C$ becomes frequency and velocity
dependent. Rybicki \& Hummer (1978) introduced a generalized Sobolev
method for non-monotonic velocity laws, where the radiation field is
found by iteration. This introduces interesting, non-local effects
into Abbott wave propagation (action at a distance). We postpone such
an analysis to a future paper, and proceed here in a simpler
fashion. Together with $\gl \sim \sqrt{|v'|}$, the opposite extreme
$\gl \sim \sqrt{ \max(v',0)}$ is treated. In the latter force, all
radiation is assumed to be absorbed at the first resonance location,
where necessarily $v'>0$. The line force according to Rybicki \&
Hummer (1978) lies in between these two extremes.

Repeating the above steps for these generalized line forces, the Euler
equation maintains its characteristic form,
 \begin{equation}
 \left(\pa_t + A \pa_z \right) {v'\over \rho}=-{g'\over \rho}\notag,
 \end{equation}
 with Abbott speed $A=v - {\dis\Gamma\over\dis\sqrt{\rho v'}}$ for
$v'>0$, and
 \begin{equation}
 A=
 \begin{cases}
  v + {\dis\Gamma\over\dis\sqrt{-\rho v'}} 
       & \text{if~$\gl\sim\sqrt{|v'|}$}\\
  v    & \text{if~$\gl\sim\sqrt{\max(v',0)}$}
 \end{cases}
 \end{equation}
 for $v'<0$. Therefore, if the velocity gradient is negative, Abbott
waves propagate downstream, with a positive (or zero) comoving frame
velocity along all solution types, whether shallow, steep, or
critical. This is peculiar, since Abbott waves appeared so far as {\it
upstream} mode. (Note that, for $\gl\sim\sqrt{\max(v',0)}$, the line
force drops out of the Euler equation if $v'<0$. Both wave modes
become ordinary sound then.) We conclude that regions with $v'<0$
cannot communicate with the wind base.

\section{Abbott wave runaway} 
\label{awrunaway}

\subsection{Method}

In the remainders, we study wave propagation in LDWs numerically,
using a standard time-explicit, Eulerian grid code (van Leer advection
on staggered grids). Non-reflecting Riemann boundary conditions for
Abbott waves are used (Feldmeier et al.~2001). As inner boundary,
$z=0.1$ is chosen to avoid negative speeds when Abbott waves leave the
mesh at the wind base: numerical artefacts may result when the $v$
characteristic changes its direction. For gravity, we assume
$g=z/(1+z^2)$, with a maximum at $z=1$. Since $g$ and $z$ are
normalized, so are $v$ and $t$. Steep solutions are of no further
interest here, since they are super-abbottic and, therefore,
numerically stable. Furthermore, since they start supersonically at
the wind base, they are unphysical. We are left with shallow winds,
which can evolve towards the critical solution by means of Abbott
waves. Shallow solutions are numerically unstable if pure outflow
boundary conditions are used. The mechanism of this runaway is not
easy to analyse, because of numerical complications in the vicinity of
the boundary, where the nature of the difference scheme changes.

To clearly separate effects of boundary conditions from wave dynamics,
we introduce controlled, explicit perturbations in the {\it middle} of
the calculational domain.

\subsection{Mechanism of the runaway}

\begin{figure}[ht]
 \begin{center}
 \leavevmode
 \epsfxsize=7.cm
 \epsffile{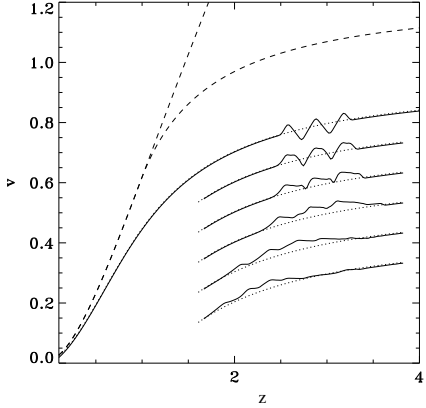}    
 \caption{Evolution of a triangular wave train in a
shallow wind (the latter shown as dotted line). For clarity, the
velocity law is plotted with a negative, vertical offset which
increases with time. The dashed lines show the shallow solution with
$m=1$ and the critical solution.\label{sawtooth}}   
 \end{center}
 \end{figure}

Figure~\ref{sawtooth} demonstrates the result derived above, that
positive and negative velocity slopes propagate in opposite
directions. The initial conditions are a shallow velocity law which is
perturbed by a triangular wave train. The subsequent evolution of this
wave train follows from the kinematics of the velocity slopes. In
strict mathematical terms, the plateaus which form in
Fig.~\ref{sawtooth} correspond to centered rarefaction waves. We
postpone such an analysis to a forthcoming paper. The essential result
from the figure is that the wind speed in the sawtooth evolves
asymmetrically, towards larger values.

 \begin{figure}[ht!!!!!]
 \begin{center}
 \leavevmode
 \epsfxsize=7.cm
 \epsffile{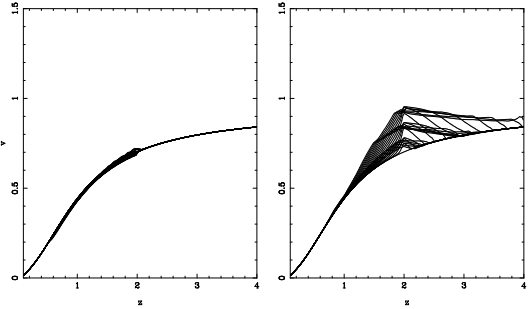}
 \caption{{\it Left panel:} stable Abbott wave
propagation along a shallow velocity law. A sinusoidal perturbation
with amplitude $S=0.04$ and period $P=1$ is applied at $\ze=2$. {\it
Right panel:} Abbott wave runaway if the amplitude is doubled to
$S=0.08$.\label{runaway}}
 \end{center}
 \end{figure} 

{\it The whole sawtooth pattern moves upstream as an Abbott wave}.
This is of a prime importance for our understanding of the observed
runaway.

Namely, if a perturbation is fed into the wind continuously over time,
the whole inner wind is eventually lifted towards larger speeds and
mass loss rates. The same is true for the outer wind, first directly
by the runaway, and second as a consequence of the accelerated inner
gas propagating outwards.

We consider a {\it coherent}, sinusoidal velocity perturbation of
period $P$ and maximum amplitude $S$, which is fed into the flow at a
fixed location $z$. The fundamental hydrodynamic variables used in the
code are $\rho$ and $\rho v$. After each time step $\delta t$,
perturbations
 \begin{equation}
 \rho v \rightarrow \rho v + \rho \,\delta v,\qquad
 \rho \rightarrow \rho(1-\delta v/|A|),
 \end{equation}
 with
 \begin{equation}
 \delta v= \delta t\, {2\pi S\over P}\cos{2\pi t\over P}
 \end{equation}
 are applied to $\rho v$ and $\rho$ on a single mesh point. The
density fluctuations follow from the continuity equation,
$\delta\rho/\rho\approx -\delta v/|A|$. For linear waves, the
observers frame Abbott speed is $A\approx -1.05$ at $\ze=2$ for
$m=0.8$.

For sufficiently small amplitudes $S$, $v'$ remains positive, and
Abbott waves propagate in a stable fashion towards the wind base. This
is shown in the left panel of Figure~\ref{runaway}, for $P=1$ and
$S=0.04$. Doubling the perturbation amplitude to $S=0.08$ implies wind
runaway towards the critical solution, as is shown in the right panel
of Figure~\ref{runaway}. The wind converges to $m=1$ everywhere (not
shown). Instead of adopting the critical, accelerating branch, the
velocity law jumps at $z>\ze$ to the decelerating branch. This is even
true for the converged, stationary solution as $t\rightarrow
\infty$. (The velocity slope is so mildly negative for $t\rightarrow
\infty$ that the wind speed is almost constant above $\ze=2$.) We add
some further remarks on this issue below. 

The runaway results from the occurence of negative velocity gradients.
During excitation phases where $v'<0$, the resulting line force
perturbations are {\it not sufficiently negative} to compensate for
positive line force perturbations during phases where $v'>0$. Net
acceleration of the wind results over a full excitation cycle is the
consequence.

 \begin{figure}[ht]
 \begin{center}
 \leavevmode
 \epsfxsize=7.cm
 \epsffile{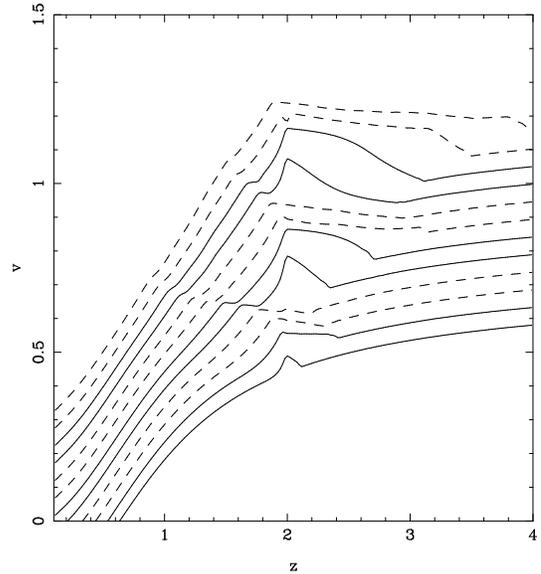}
 \caption{The runaway time series from
Fig.~\ref{runaway}, shown with a constant vertical displacement
between snapshots. Dashed lines show phases where negative velocity
perturbations are applied at $\ze$, leading to self-annihilating
slopes.\label{offset}}
 \end{center}
 \end{figure}  

Figure~\ref{offset} shows again the runaway time series of
Fig.~\ref{runaway}, with subsequent snapshots displaced vertically for
clarity. During the negative perturbation half-cycle, $-\delta
v(\ze)$, negative velocity slopes propagate outwards from below $\ze$,
and merge with positive slopes propagating inwards from above $\ze$.
The merging slopes mutually annihilate. After a time $P/2$, the
velocity law is left largely unaltered in presence of the
perturbation. This causes the dense spacing of curves in
Figure~\ref{runaway}, especially at $z>\ze$, once every perturbation
cycle. On this rather flat velocity law, a positive perturbation
$+\delta v$ is added during the next half-cycle. Here, inner, positive
slopes propagate inwards and separate from outer, negative slopes
which propagate outwards. Obviously, runaway is caused by positive
perturbations, but its deeper origin is that negative perturbations
are self-annihilating, and cannot balance positive ones.

The runaway terminates when $\ze$ comes to lie on the critical
solution, and Abbott waves can no longer propagate inwards.

The velocity gradient at $z>\ze$ is then still negative, and in most
of our simulations remains negative at all times: at $\ze$, the wind
jumps to the decelerating branch with $m=1$, which causes a kink in
$v(z)$. Runaway perturbations above $\ze$ would be required to
establish a critical, accelerating velocity law in the outer wind. For
certain combinations of model parameters, we find instead a critical
solution with $v'>0$ over the whole mesh. The near-plateau above $\ze$
evolves then towards larger speeds, and the velocity kink propagates
outwards, eventually leaving the mesh. We believe that this is an
artefact caused by boundary-induced, numerical runaway (Feldmeier \&
Shosman 2001). The latter can even occur when non-reflecting Abbott
boundary conditions are used, via {\it non-WKB} (standing?) waves; our
Riemann boundary conditions were formulated to annihilate WKB waves
only. We prefer the situation $v'<0$ for $z>\ze$ over any numerical
runaway which would assist the present, physical runaway in reaching
the critical solution. A future analysis of outer boundary conditions
has to clarify this issue of the outer wind velocity law.

\begin{figure}[ht]
 \begin{center}
 \leavevmode
 \epsfxsize=7.cm
 \epsffile{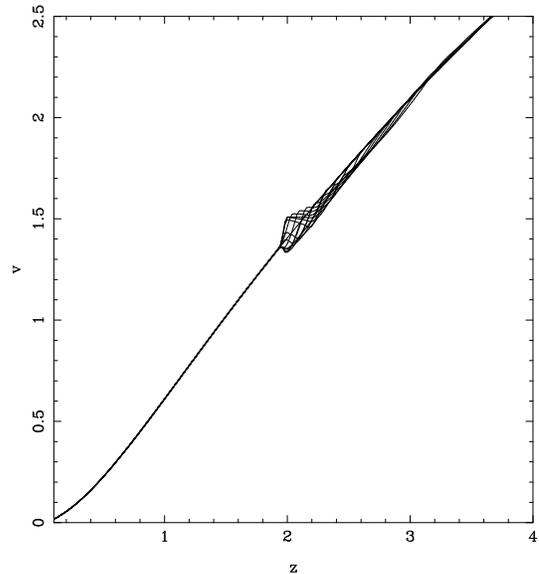}
 \caption{The critical, CAK solution is stable with
respect to Abbott waves excited above the critical point. Perturbation
phases with positive and negative $v'$ combine to an outward
propagating, harmonic Abbott wave.\label{critstable}}
 \end{center}
 \end{figure}

We can still test the stability of the full, critical solution, as is
done in Figure~\ref{critstable}. Above the critical point,
perturbation phases with $v'>0$ and $v'<0$ combine here, and propagate
outwards as a smooth, marginally stable Abbott wave.

\subsection{Stationary overloaded winds}
\label{overloading1}

So far, we assumed that wind perturbations are located above the
critical point. We consider now the opposite case, $\ze < z\cri$.

\begin{figure}[ht]
 \begin{center}
 \leavevmode
 \epsfxsize=7.cm
 \epsffile{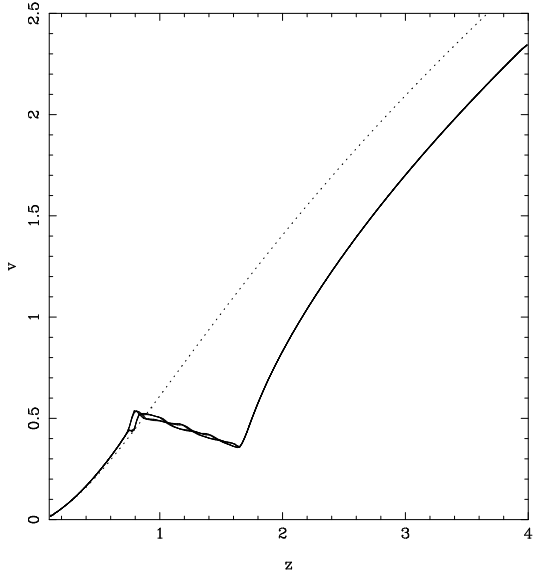}
 \caption{Velocity law for a wind with harmonic
perturbation at $\ze=0.8$, below the critical point $z\cri=1$. The
wind converges to a stationary, overloaded solution. The dotted line
shows the initial conditions, the critical CAK solution.\label{overloaded}}
 \end{center}
 \end{figure}

Figure~\ref{overloaded} shows the wind velocity law resulting from a
perturbation location $\ze=0.8$ on the critical solution. A period
$P=0.3$ was chosen. A stationary wind with a broad deceleration
region, $v'<0$, develops. Abbott waves propagate {\it outwards} from
$\ze$ through the decelerating wind. Since the CAK critical point,
which is the bottleneck of the flow, lies in the deceleration regime,
the present solution should be overloaded. Indeed, the mass loss rate
is found to be $m=1.05$.

This result is readily understood. The flow at $\ze$ is still
sub-abbottic, because $\ze<z\cri$. Runaway to larger $v$ occurs until
inward propagation from $\ze$ becomes impossible. The velocity at
$z<\ze$ is then everywhere {\it larger} than the critical speed. The
runaway stops when the inward Abbott speed becomes zero in the
observers frame. We are interested in a stationary solution, hence,
$A=0$ in (\ref{staabbspee}), implying $mw'=g\cri$. The square root in
(\ref{cakwindsol}) has to vanish then, hence $m=g_c/g$.

So far, we restricted ourselves to $g=g_c$ and $m=1$. But critical
points can also occur along overloaded solutions. In this case,
$m=g\cri/g(z\cri)>1$, and the square root in (\ref{cakwindsol})
becomes imaginary at $z>\ze$. Remember that $g\cri$ refers to gravity
at the critical point of the {\it critical} solution, which is the
maximum of $g$. Hence, $g(z\cri)<g\cri$ for any other possible
critical point. The perturbation site $\ze$ stops communicating with
the wind base only when it becomes a critical point itself. At $\ze$,
the overloaded wind jumps to the decelerating branch, resulting in a
kink in the velocity law. This kink has all the attributes of a
critical point. Especially, as an Abbott wave barrier, it shuts off
communication with the wind base.

It is easily seen that a critical point at $z<\ze$ would imply that
$\ze$ is already super-abbottic, which cannot happen during runaway.
Since $g(z)$ is a monotonically growing function with a maximum $g_c$,
the perturbation site $\ze<z\cri$ is the first one that becomes
critical to Abbott waves on the evolving overloaded solution. We
emphasize that the amount of overloading is solely determined by
$\ze$.

At some height above the gravity maximum, the decelerating solution
jumps back to the accelerating branch, giving a second kink: the wind
has overcome the bottleneck, and starts to accelerate again. In the
present wind model, the (imaginary, overloaded) solution becomes real
again at $z=1/\ze$ ($\ze$ being the location where the solution
becomes first imaginary). For $\ze=0.8$, the overloading is $m=1.025$
and the wind can start accelerating again at $z=1.25$. In the
simulation however, $m=1.05$, and the second kink occurs at $z=1.6$.
The discrepance in $m$ can be attributed to mesh resolution, which
blurrs out $\ze$. This leaves still unanswered why the wind `waits' so
long before it starts accelerating again.

Furthermore, starting with a shallow wind as initial conditions
instead of a critical wind, the simulation converges again to an
overloaded solution. However, the jump from the decelerating back to
the accelerating branch does not terminate on the steep, but already
on the shallow overloaded solution with $m=1.05$. This is a numerical
artefact caused by non-reflecting, outer boundary conditions, which
try to maintain shallow solutions. As above, we prefer this situation
over any boundary-induced, numerical runaway.

A third kink may occur. For realistic radiative fluxes above accretion
disks, we find that at still larger distances from the disk, the line
force drops below absolute gravity for {\it all} solution types
(Feldmeier \& Shlosman 1999). Another jump to the decelerating branch
is unavoidable. Since the local wind speed is much larger than the
local escape speed, the velocity law stays essentially flat above this
kink.

\subsection{Time-dependent overloaded winds}
\label{overloading2}

\begin{figure}[ht]
 \begin{center}
 \leavevmode
 \epsfxsize=7.cm
 \epsffile{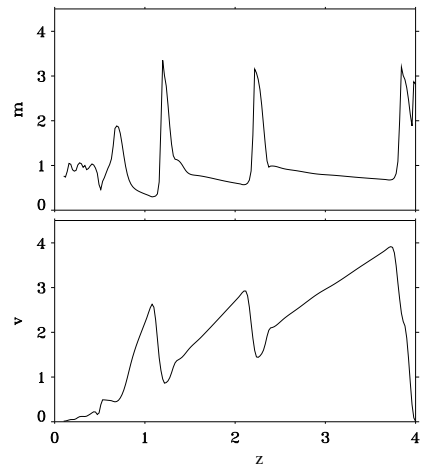}
 \caption{Velocity $v$ and mass flux $m$ of a
wind perturbed at $\ze=0.5$. During each perturbation cycle, gas falls
back towards the photosphere and collides with upwards streaming
gas. Shocks form, and propagate outwards.\label{shocksandshells}}
 \end{center}
 \end{figure}

Already a minor increase in the mass loss rate causes a broad
deceleration region. This is a consequence of $g(z)$ having a broad
maximum at $z_c$. Integrating $w'(z)=vv'$ numerically, we find that
$v(1/\ze)=0$ for $\ze=0.660$. If $\ze$ is still smaller, the gas
starts to fall back towards the photosphere before it reaches $1/\ze$,
and collides with upwards streaming gas. A stationary solution is no
longer possible. Figure~\ref{shocksandshells} shows, for $\ze=0.5$,
that an outward propagating sequence of shocks and dense shells
forms. The shock spacing is determined by the perturbation period.

As a technical comment, we add that the latter simulation tends to
develop extremely strong rarefactions and, correspondingly, very large
velocities. The latter cause the Courant time step to approach
zero. This is a well-known artefact of the power law form of the CAK
line force, $\gl\propto (v'/\rho)^\alpha$, implying $\gl\rightarrow
\infty$ as $\rho\rightarrow 0$. The true line force reaches instead a
finite maximum in a flow which is optically thin even in very strong
lines (Abbott 1982). This is achieved by truncating the line
distribution function (Owocki, Castor, \& Rybicki 1988). Using a
simpler approach which suffices for the present purposes, we simply
truncate $m$ at large values.

Even for $\ze=0.660$ implying negative speeds, overloading is small,
only 9\% above the critical CAK value ($m=1.088$). This shows that the
CAK mass loss rate is a significant upper limit for mass loss rates in
LDWs, with or without mass overloading. If overloading occurs, it
should be detectable only via broad, decelerating flow regions.

It is important to distinguish the present runaway from the well-known
line-driven instability (Lucy \& Solomon 1970; Owocki \& Rybicki 1984;
Lucy 1984), which leads to strong shocks and dense, narrow shells in
the wind. For perturbations longer than the Sobolev scale, the
instability can be understood as a linear process in second-order
Sobolev approximation, including curvature terms $v''$. By contrast,
runaway occurs already in first-order Sobolev approximation, yet,
finite perturbations are required to achieve $v'<0$.

\subsection{Triggering perturbations}

{\it Amplitudes.} Still, even perturbations with small amplitudes can
lead to runaway, if their wavelength is sufficiently short to cause
negative $v'$. One expects, however, that dissipative processes like
heat conduction prevent runaway for short wavelength perturbations.

{\it Coherence.} In a hydrodynamic instability, a feedback cycle
amplifies small initial perturbations. Even perturbations of short
duration trigger unstable growth. By contrast, the present wind
runaway requires a continuously maintained perturbation seed. No
feedback occurs. The runaway is a pure wave feature caused by the
peculiar $v'$-asymmetry of the line force.

Perturbations of short duration lead to a localized runaway. The wind
is shifted towards larger $v$ and $m$ within a small $z$ interval. If
the perturbation ceases, so does the runaway. The region of increased
$v$ and $m$ propagates upstream as an Abbott wave, and eventually
lifts the wind base to a {\it stable} shallow solution with slightly
higher $m$. Successive Abbott waves lift the wind until the critical
or an overloaded solution is reached.

{\it Surrounding medium.} One expects that mismatches at the outer
boundary, where the wind propagates into a medium of given properties,
create perturbations which can trigger runaway. This is supported by
the fact that at large $z$ the velocity law is almost flat, and even
small-amplitude perturbations can cause $v'<0$. The Abbott speed is
large at large $z$, since $A\sim 1/\sqrt{v'}$, hence perturbations
propagate inwards quickly. We performed tests with the outer boundary
at $z=40$. Inherent interpolation errors are sufficient then to set
off boundary runaway.

{\it Other sources.} Inner wind perturbations could occur due to
prevalent shocks from the line driven instability (Owocki et
al.~1988). The critical point lies usually close to the sonic point
(Owocki \& Puls 1999; Feldmeier \& Shlosman 1999), and it is therefore
not easy to contemplate strong shocks below the critical point. This
means that perturbations at large and not at small $z$ dominate the
runaway, and drive the wind to a critical solution. Hence, overloaded
winds formed by internally-generated perturbations should be rare.

 \subsection{Critical points and mass loss rate}

We discuss here some general issues related to critical points and
mass loss rates in different types of stellar winds.

{\it Holzer's wind laws.} The idea that upstream, inwards propagating
waves adapt the wind base to outer (boundary) conditions is
fundamental to solar wind theory. From this grew the recognition that
an outward force which is applied above the sonic (critical) point
does not affect the mass loss rate, but only accelerates the flow;
whereas a force applied below the sonic point increases the mass loss
rate, but has a vanishing effect on terminal wind speeds (Leer \&
Holzer 1980; Holzer 1987). These `wind laws' have proven to be of
interest for cases far beyond the coronal winds for which they were
first applied. We refer to Lamers \& Cassinelli (1999) for a detailed
discussion.

Most relevant to us is the case of a force applied in the subcritical
wind region, causing enhanced mass loss. For coronal winds, the
subsonic region has essentially a barometric density stratification.
Any extra force assisting the pressure gradient helps to establish a
larger scale height. From the continuity equation, a shallower
velocity gradient results, though the terminal speed is hardly
affected. The corresponding situation, analyzed by us for LDWs, is
even simpler. No outward force occurs besides line driving. Our choice
of zero sound speed emphasizes that the barometric stratification
plays no role for mass loss runaway or for establishing an overloaded
solution. Overloaded LDW solutions have a {\it steeper} velocity
gradient than the critical solution, as can be seen from
eq.~(\ref{eulerstatsolu}). Physically, a steeper velocity gradient is
required to create a generalized critical point below the CAK critical
point. In LDWs, a critical {\it acceleration} $w'=vv'$, not a critical
speed $v$ is adopted at the critical point, and prevents further
Abbott wave runaway.

{\it Physical relevance of critical points.} The physical relevance of
the CAK critical point was questioned by Lucy (1975, 1998), arguing
that it may be an artifact of Sobolev approximation. Holzer (1987,
p.~296) doubts that critical points, i.e., singularities of the flow
equations, generally coincide with those points beyond which
``information relevant to the acceleration of the wind'' can no longer
be transported upstream. We have tried in the present paper to
re-establish a more traditional viewpoint (Courant \& Friedrichs 1948;
Courant \& Hilbert 1968), namely that a critical point as a
mathematical singularity leaves certain derivatives of flow quantities
undetermined. This is only possible along characteristics, which are
the space-time trajectories of the specific {\it waves} in the
problem. This establishes the transonic (or trans-abbottic, etc)
nature of the critical point. Similarly, one can argue: the critical
point is a singularity because different solution branches cross
there. Hence, by passing continuously from one branch to another
(actually: by staying on a certain branch), higher-order or weak
discontinuities appear in the flow properties; for LDWs, in
$v''$. Again, weak discontinuities propagate at characteristic wave
speed, making the critical point transonic.

{\it Mass loss rate as an eigenvalue.} We remind the reader of a deep
difference between coronal and line-driven winds. For coronal winds,
the mass loss rate is a free parameter within wide margins, and is
determined by the base density. For LDWs, on the other hand, the mass
loss rate is a unique, discrete eigenvalue (CAK). This is a
consequence of the line {\it acceleration} depending on $\rho$, see
Lamers \& Cassinelli (1999).

{\it Abbott speed and speed of light.} In the literature, the
propagation speed of radiative waves is occasionally identified with
the speed of light. This is not true in general: the Abbott speed and,
in magnetized flows, the Alfv\'en speed are smaller than the speed of
light. To make the point totally obvious, note that the diffusion
speed of photons through stellar interiors is much smaller than the
speed of light. The basic cause is here the same as for Abbott waves:
optical depths larger than unity.

\subsection{Deep-seated X-rays, infall, and mass-over-loading}

Besides adapting the wind base to outer flow conditions, we find that
Abbott waves can even lead to mass inflow. This seems like a novel
feature of these waves and it leads to interesting observational
consequences, like explaining the formation of hard X-rays anomalously
close to the surfaces of hot stars. Formation depths of $r\approx 1.1
\,R_\ast$ are deduced from X-ray emission lines observed with the {\sc
Chandra} satellite (for $\zeta$~Ori: Waldron \& Cassinelli 2000; for
$\zeta$~Pup: Cassinelli et al.~2001). The favored model for X-ray
emission from hot stars is via strong shocks (Lucy 1982) from the
de-shadowing instability (Lucy \& Solomon 1970; Owocki \& Rybicki
1984). The shocks become strongest in collisions of fast clouds with
dense gas shells (Feldmeier, Puls, \& Pauldrach 1997). From
theoretical arguments (line drag effect: Lucy 1984) and numerical
simulations it appears that this shock scenario cannot explain X-rays
originating from the small heights mentioned above.

Howk et al.~(2000) suggest gas infall as possible origin of X-rays
from near the photosphere. They consider a ballistic model of stalled
wind clouds falling back towards the star, to explain the hard X-rays
observed from $\tau$~Sco (Cassinelli et al.~1994). At any instant,
$\approx 10\%$ of the mass flux from the star resides in numerous,
$\ge 1000$, clouds which barely reach distances of $r=2\,R_\ast$
before falling back inwards. The overall resemblance of this cloud
model with time-dependent wind overloading as shown in
Fig.~\ref{shocksandshells} is striking. In the latter case, the
overloading is sufficiently mild that upstreaming gas can push
downfalling clumps outwards (cf.~the drag force of ambient wind gas in
Howk et al.). Stronger overloading is achieved by shifting the Abbott
wave source further inwards, but cannot be addressed using our simple
numerical approach. Clearly, two-dimensional simulations are required
to model true infall. A systematic study relating the Howk et
al.~approach with ours is in preparation.

Note that the critical point in stellar wind models accounting for the
finite cone effect lies at $r\le 1.1 \,R_\ast$ (Pauldrach et
al.~1986). An overloaded wind starts to decelerate below the critical
point, and may already reach negative or infall speeds at similarly
small heights. This shows the relevance of overloading in
understanding the {\sc Chandra} observations mentioned above. The
topic of stalling and backfalling gas has aquired some general
attention in recent papers on LDWs, and is discussed in different
contexts by Friend \& Abbott (1986), Poe et al.~(1990), Koninx (1992),
Proga, Stone, \& Drew (1998), and Porter \& Skouza (1999).

We finally mention another idea related to overloaded flows, which has
played some role in sharpening our understanding of coronal winds.
Cannon \& Thomas (1977) and Thomas (1982) challenged Parker's (1958,
1960) theory of the solar wind. They postulated an outwards directed,
sub-photospheric mass flux. In analogy with a Laval nozzle (e.g.,
Chapman 2000, Chapter~7) into which a too strong mass or energy influx
is fed (Cannon \& Thomas 1977 use the terminology `imperfect nozzle'
instead of `overloading'), the solar outflow should choke, creating
shocks below the nozzle throat, or the critical point of the smooth
(perfect nozzle) flow. The shocks are responsible for heating the
chromosphere and corona, making the latter a consequence, not the
origin of outflow from Sun. This model was ruled out by Parker (1981)
and Wolfson \& Holzer (1982).

Still, it leads over to another interesting difference between coronal
and Laval nozzle flows on the one side and LDWs on the other: the
former two do not allow for continuous, overloaded solutions; instead,
overloaded flows have shocks in the $vr$ plane where the critical
point resides. By contrast, overloaded LDWs show jumps in the critical
$vv'$-$r$ plane, i.e., jumps in the wind acceleration which correspond
to {\it kinks} in a continuous velocity law.
 
\section{Summary}
\label{summary}

We have studied the stability of shallow and steep solutions for
accretion disk and stellar line driven winds. This was done by
introducing flow perturbations into the wind at a fixed height. For
sufficiently large perturbation amplitudes, negative velocity
gradients occur in the wind and cause runaway towards the critical
solution. The origin of this runaway is an asymmetry in the line
force: negative velocity gradients cause a force decrease which cannot
balance the force increase during phases where the velocity law gets
steepened. Net acceleration results over a full perturbation cycle.

A new type of waves, termed Abbott waves, are exited by the
perturbations. They provide a communication channel between different
parts of the wind and define an additional critical point in the flow,
downstream from the sonic point. Shallow solutions are subcritical
everywhere, steep solutions are supercritical. Along shallow
solutions, Abbott waves propagate upstream towards the photosphere at
a high speed as a radiative mode, and creep downstream at a fraction
of the sonic speed as an acoustic mode. The critical solution is the
one that switches continuously from the shallow to the steep branch at
the critical point.

Inward propagating Abbott waves turn the local, asymmetric response to
velocity perturbations into a global runaway towards the critical
solution. The converged, steady wind solution depends on the location
of seed perturbations. Three spatial domains can be distinguished. (i)
If perturbations are prevalent in outer wind regions, above the
critical point, the wind settles on the critical solution. (ii) If
perturbations occur just below the critical point, runaway proceeds to
a stationary, so-called overloaded solution. Such a flow is
characterized by a broad deceleration domain. The line force cannot
balance gravity in a vicinity of the critical point (the bottleneck of
the flow) because of a supercritical mass loss rate. (iii)
Perturbations close to the photosphere result in the wind being
decelerated to negative speeds. Gas falls back towards the
photosphere, and collides with upwards streaming gas. A
time-dependent, overloaded wind results with a train of shocks and
shells propagating outwards.

The runaway mechanism discussed here depends solely on the asymmetry
of the line force with respect to velocity perturbations which cause
local flow deceleration, $dv/dr<0$. It should, therefore, be a robust
feature and not depend on Sobolev approximation.

\acknowledgements

We thank Mark Bottorff, Wolf-Rainer Hamann, Colin Norman, Stan Owocki,
and Joachim Puls for illuminating discussions. We thank the referee,
Joe Cassinelli, for suggestions towards a more detailed discussion of
critical points and gas infall. This work was supported in part by
NASA grants NAG 5-10823, NAG5-3841, WKU-522762-98-6 and HST
GO-08123.01-97A to I.S., which are gratefully acknowledged.


\begin{references}

\reference{} Abbott, D.C. 1980, ApJ, 242, 1183
\reference{} Abbott, D.C. 1982, ApJ, 259, 282
\reference{} Cannon, C.J. \& Thomas, R.N. 1977, ApJ, 211, 910
\reference{} Cassinelli, J.P., Cohen, D.H., MacFarlane, J.J., Sanders, W.T. 
             \& Welsh, B.Y. 1994, ApJ, 421, 705
\reference{} Cassinelli, J.P., Miller, N.A., Waldron, W.L., MacFarlane, J.J.
             \& Cohen, D.H. 2001, ApJ, 554, L55
\reference{} Castor, J.I., Abbott, D.C. \& Klein, R.I. 1975, ApJ, 195, 157 (CAK)
\reference{} Chapman, C.J. 2000, High Speed Flow (Cambridge: Cambridge Univ. Press)
\reference{} Courant, R. \& Friedrichs, K.O. 1948, Supersonic Flow and Shock Waves
             (New York: Interscience Publishers)
\reference{} Courant, R. \& Hilbert, D. 1968, Methoden der Mathematischen Physik
             (Berlin: Springer)
\reference{} Feldmeier, A., Puls, J. \& Pauldrach, A. 1997, A\&A, 322, 878
\reference{} Feldmeier, A. \& Shlosman, I. 1999, ApJ, 526, 344
\reference{} Feldmeier, A. \& Shlosman, I. 2000, ApJ, 532, L125
\reference{} Feldmeier, A., Shlosman, I. \& Hamann, W.-R. 2001, ApJ, submitted
\reference{} Friend, D.B. \& Abbott, D.C. 1986, ApJ, 311, 701
\reference{} Holzer, T.E. 1987, IAU Symp. \#122, Circumstellar Matter
      (Dordrecht: Reidel), 289  
\reference{} Howk, J.C., Cassinelli, J.P., Bjorkman, J.E. \&       
      Lamers, H.J.G.L.M. 2000, ApJ, 534, 348 
\reference{} Koninx, J.-P. 1992, Ph. D. thesis, Utrecht Univ. 
\reference{} Lamb, H. 1932, Hydrodynamics (New York: Dover) 
\reference{} Lamers, H.J.G.L.M. \& Cassinelli, J.P. 1999, Introduction
       to Stellar Winds (Cambridge: Cambridge Univ. Press) 
\reference{} Leer, E. \& Holzer, T.E. 1980, JGR, 85, 4681 
\reference{} Lucy, L.B. 1975, Memoires Societe Royale des Sciences de
      Liege, 8, 359 
\reference{} Lucy, L.B. 1982, ApJ, 255, 286
\reference{} Lucy, L.B. 1984, ApJ, 284, 351
\reference{} Lucy, L.B. 1998, in Cyclical Variability in Stellar Winds, ed.
             L. Kaper \& A.W. Fullerton (Berlin: Springer), 16
\reference{} Lucy, L.B. \& Solomon, P.M. 1970, ApJ, 159, 879
\reference{} Owocki, S.P. \& Rybicki, G.B. 1984, ApJ, 284, 337
\reference{} Owocki, S.P. \& Rybicki, G.B. 1986, ApJ, 309, 127
\reference{} Owocki, S.P., Castor, J.I. \& Rybicki, G.B. 1988, ApJ, 335, 914
\reference{} Owocki, S.P. \& Puls, J. 1999, ApJ, 510, 355
\reference{} Parker, E.N. 1958, ApJ, 128, 644
\reference{} Parker, E.N. 1960, ApJ, 132, 175
\reference{} Parker, E.N. 1981, ApJ, 251, 266
\reference{} Pauldrach, A., Puls, J. \& Kudritzki, R.P. 1986, A\&A, 164, 86
\reference{} Poe, C.H., Owocki, S.P. \& Castor, J.I. 1990, ApJ, 358, 199
\reference{} Porter, J.M. \& Skouza, B.A. 1999, A\&A, 344, 205
\reference{} Proga, D., Stone, J.M. \& Drew, J.E. 1998, MNRAS, 295, 595
\reference{} Rybicki, G.B. \& Hummer, D.G. 1978, ApJ, 219, 654
\reference{} Thomas, R.N. 1982, ApJ, 263, 870
\reference{} Waldron, W.L. \& Cassinelli, J.P. 2000, ApJ, 548, L45
\reference{} Wolfson C.J. \& Holzer, T.E. 1982, ApJ, 255, 610
\end{references}
\end{document}